\documentclass[aps,pra,twocolumn,superscriptaddress,showpacs]{revtex4}%
\usepackage{graphics}
\usepackage{amsmath}
\usepackage{graphicx}
\usepackage{amsfonts}
\usepackage{amssymb}
\newtheorem{theorem}{Theorem}

\newtheorem{corollary}[theorem]{Corollary}

\begin{document}
\title{A limit formula for the quantum fidelity}
\author{Gaetana Spedalieri}
\email{gae.spedalieri@york.ac.uk}
\affiliation{Department of Computer Science, University of York, York YO10 5GH, United Kingdom}
\author{Christian Weedbrook}
\affiliation{Center for Quantum Information and Quantum Control, Department of Electrical
and Computer Engineering and Department of Physics, University of Toronto,
Toronto, M5S 3G4, Canada}
\author{Stefano Pirandola}
\affiliation{Department of Computer Science, University of York, York YO10 5GH, United Kingdom}
\date{\today }

\begin{abstract}
Quantum fidelity is a central tool in quantum information, quantifying how
much two quantum states are similar. Here we propose a limit formula for the
quantum fidelity between a mixed state and a pure state. As an example of an
application, we apply this formula to the case of multimode Gaussian states,
achieving a simple expression in terms of their first and second-order
statistical moments.

\end{abstract}

\author{}
\maketitle

\section{Introduction}

Very often the performance of a quantum information protocol is measured in
terms of similarity between two states. This happens both in dicrete (qubit)
quantum information~\cite{NielsenBook} and continuous-variable quantum
information~\cite{BraREV,BraREV2}, for instance, with Gaussian
states~\cite{RMP,GaussSTATES}.

One of the most well known examples is that of quantum teleportation between
two stations, where the perfect execution of the protocol corresponds to
having an output state at the receiver's station which is equal to the input
state originally processed at the sender's
station~\cite{CVtelepo,Bra98,RalphTELE,TeleNET,PirTeleOPMecc,Barlett2003,Pirgames,Pir2005,teleREV,Sherson}%
. Another important scenario is that of quantum
cloning~\cite{Buzek,lin,Cerf1,braclone,fiu}, where an unknown input quantum
state is transformed into 2 or more clones. Because of the no-cloning theorem,
the output clones cannot be identical to the input
state~\cite{Noclone1,Noclone2}. As a result, a measure of similarity between
input state and output clones is fundamental in order to quantify the
performance of a quantum cloning machine.

In these kinds of protocols, the typical measure of similarity between two
quantum states is their fidelity. Quantum fidelity was introduced and
characterized in two seminal papers by Uhlmann~\cite{Uhlmann} and
Jozsa~\cite{Jozsa}. Its general definition can be regarded as an extension of
the wavefunction overlap to generally mixed quantum states. Despite its
usefulness, simple closed formulas are not always easy to derive. For
instance, we know a closed formula for the fidelity between two single-mode
Gaussian states~\cite{fidG1,fidG2,fidG3,fidG4}, but a simple analytical result
is still missing in the case of two arbitrary multimode Gaussian states.

Quantum fidelity also plays a central role in quantum hypothesis testing,
where the basic problem is the discrimination between two equiprobable quantum
states of a system by means of an optimal measurement. In this framework,
quantum fidelity has been related to various other quantities such as the
Helstrom bound~\cite{Helstrom,Fuchs,FuchsThesis} and the quantum Chernoff
bound~\cite{QCbound,MinkoPRA}, which provide a direct quantification of the
minimum error probability affecting the state discrimination.

In this paper, we start from these connections to derive a new formula for the
quantum fidelity between a generally mixed state and a pure state. This
formula is expressed in the form of a limit and involves a generalized overlap
between the two quantum states. This is the same kind of overlap which
intervenes in the definition of the quantum Chernoff bound.

As an example of an application of this formula, we consider the bosonic
setting and the case of multimode Gaussian states. Here we first introduce the
notion of symplectic action, which enables us to simplify the symplectic
manipulations of the second order moments. Then, by elaborating a result from
Ref.~\cite{MinkoPRA}, we derive a simple analytical expression for the
fidelity between a mixed and a pure Gaussian state in terms of their first and
second order moments.

The paper is organized as follows. In Sec.~\ref{Sec1}\ we provide a brief
review of the basic facts regarding quantum fidelity and its connections with
the various bounds used in quantum hypothesis testing. In Sec.~\ref{Sec2}, we
derive the limit formula for the quantum fidelity. Then, in Sec.~\ref{Sec3},
we consider bosonic continuous-variable systems. After a brief review of the
basic notions on Gaussian states, we introduce the symplectic action and we
derive the formula of the fidelity for Gaussian states. Finally,
Sec.~\ref{Sec4} is for conclusions.

\section{General notions on quantum fidelity\label{Sec1}}

Consider a quantum system with separable Hilbert space $\mathcal{H}$. In
general, the states of this system are described by density operators
$\rho:\mathcal{H}\rightarrow\mathcal{H}$ forming a corresponding state
space\ $\mathcal{D}(\mathcal{H})$. Given two arbitrary states, $\rho$ and
$\sigma$, their similarity can be quantified by the Uhlmann-Jozsa
fidelity~\cite{Uhlmann,Jozsa,BuresFID}%
\begin{equation}
F(\rho,\sigma):=\left(  \mathrm{Tr}\sqrt{\sqrt{\rho}\sigma\sqrt{\rho}}\right)
^{2}~.
\end{equation}
This is a positive number $0\leq F\leq1$, where $F=1$ corresponds to identical
states, and $F=0$ corresponds to orthogonal states, i.e., density operators
with orthogonal supports in $\mathcal{D}(\mathcal{H})$. In the case where one
of the states is pure $\sigma=\left\vert \varphi\right\rangle \left\langle
\varphi\right\vert $, the fidelity assumes the sandwich expression%
\begin{equation}
F(\rho,\left\vert \varphi\right\rangle )=\left\langle \varphi\right\vert
\rho\left\vert \varphi\right\rangle ~,
\end{equation}
which becomes the overlap
\begin{equation}
F(\left\vert \psi\right\rangle ,\left\vert \varphi\right\rangle )=\left\vert
\left\langle \psi\right.  \left\vert \varphi\right\rangle \right\vert ^{2}~,
\end{equation}
if also the other state is pure $\rho=\left\vert \psi\right\rangle
\left\langle \psi\right\vert $.

Most of the properties of the quantum fidelity can be derived from Uhlmann's
theorem which states that
\begin{equation}
F(\rho,\sigma)=\max_{\left\vert \varphi_{\sigma}\right\rangle }\left\vert
\left\langle \varphi_{\rho}\right.  \left\vert \varphi_{\sigma}\right\rangle
\right\vert ^{2}~,
\end{equation}
where $\left\vert \varphi_{\rho}\right\rangle $ and $\left\vert \varphi
_{\sigma}\right\rangle $ are purifications of $\rho$ and $\sigma$,
respectively. For instance, immediate consequences of this theorem are the
positive range $0\leq F\leq1$, the symmetry property $F(\rho,\sigma
)=F(\sigma,\rho)$, and the invariance $F(\rho,\sigma)=F(U\rho U^{\dagger
},U\sigma U^{\dagger})$ under a generic unitary $U$.

Despite being a measure of similarity between two quantum states, the quantum
fidelity is not properly a metric in the state space $\mathcal{D}%
(\mathcal{H}).$ In fact, by definition, a metric in $\mathcal{D}(\mathcal{H})$
is a map $D:(\rho,\sigma)\rightarrow\mathbb{R}$ with the following properties:

\begin{description}
\item[(i)] Positive definiteness, i.e., $D(\rho,\sigma)\geq
0~(=0\Leftrightarrow\sigma=\rho)$;

\item[(ii)] Symmetry, i.e., $\ D(\rho,\sigma)=D(\sigma,\rho)$;

\item[(iii)] Subadditivity or triangle inequality, i.e., $D(\rho,\gamma)\leq
D(\rho,\sigma)+D(\sigma,\gamma)$, for any triplet $\rho$, $\sigma$ and
$\gamma$.
\end{description}

\noindent In this list, the fidelity fails both the first property (since
$F(\rho,\rho)=1$) and the subadditivity.

Even if it is not a metric by itself, we can easily connect the quantum
fidelity to a metric in $\mathcal{D}(\mathcal{H})$. For instance, we can
consider the Bures' distance~\cite{FuchsThesis}%
\begin{equation}
D_{B}(\rho,\sigma)=\sqrt{2-2\sqrt{F(\rho,\sigma)}}~,
\end{equation}
or the angular distance~\cite{NielsenBook}
\begin{equation}
D_{A}(\rho,\sigma)=\text{\textrm{Arccos}}\sqrt{F(\rho,\sigma)}~.
\end{equation}
Most importantly, the quantum fidelity can be connected with the trace
distance, which is the standard metric adopted in quantum information for its
direct interpretation in quantum hypothesis testing.

Given two quantum states, $\rho$ and $\sigma$, their trace distance is defined
as~\cite{NielsenBook,Kolmo}%
\begin{equation}
D(\rho,\sigma)=\frac{1}{2}\left\Vert \rho-\sigma\right\Vert _{1}~,
\end{equation}
where
\begin{equation}
\left\Vert O\right\Vert _{1}:=\mathrm{Tr}\left\vert O\right\vert
=\mathrm{Tr}\sqrt{O^{\dagger}O}%
\end{equation}
is the trace norm of an arbitrary trace-class operator $O$~\cite{TraceCLASS}.
The trace distance ranges in the positive interval $[0,1]$, with $D=0$ for
identical states and $D=1$ for orthogonal states. $D(\rho,\sigma)$ determines
the error probability which affects the discrimination of the two states,
$\rho$ and $\sigma$, by means of an optimal quantum measurement. Suppose that
a system is prepared in one of two equiprobable states, $\rho$ and $\sigma$,
then the optimal positive operator valued measure (POVM) provides the correct
answer with an error probability given by the Helstrom bound~\cite{Helstrom}%
\begin{equation}
P_{err}=\frac{1-D}{2}~.
\end{equation}
According to Ref.~\cite{Fuchs}, we can use the trace distance to write the
following upper bound for the fidelity%
\begin{equation}
F\leq1-D^{2}~.
\end{equation}
Then, according to Ref.~\cite{NielsenBook} we can also write the lower bound%
\begin{equation}
(1-D)^{2}\leq F~.
\end{equation}
In particular, if one of the states is pure $\sigma=\left\vert \varphi
\right\rangle \left\langle \varphi\right\vert $ we have the tighter lower
bound~\cite{NielsenBook}%
\begin{equation}
1-D\leq F~.
\end{equation}
Finally if both the states are pure, $\rho=\left\vert \psi\right\rangle
\left\langle \psi\right\vert $ and $\sigma=\left\vert \varphi\right\rangle
\left\langle \varphi\right\vert $, then we have the equality~\cite{Helstrom}%
\begin{equation}
F=1-D^{2}~.
\end{equation}

Besides the trace distance and the Helstrom bound, the quantum fidelity
possesses important relations with other crucial quantities in quantum
hypothesis testing: the quantum Chernoff bound~\cite{QCbound} and the quantum
Battacharyya bound~\cite{MinkoPRA}. Let us consider the quantity%
\begin{equation}
C_{s}(\rho,\sigma):=\mathrm{Tr}(\rho^{s}\sigma^{1-s})\leq1~, \label{Cs_AUX}%
\end{equation}
which represents a generalized $s-$overlap between the two states $\rho$ and
$\sigma$. Using Eq.~(\ref{Cs_AUX}), we can define the Chernoff term%
\begin{equation}
C(\rho,\sigma):=\underset{s\in(0,1)}{\mathrm{\inf}}C_{s}(\rho,\sigma)~,
\label{C_inf_rho_sigma}%
\end{equation}
and the Battacharyya term%
\begin{equation}
B(\rho,\sigma):=C_{1/2}(\rho,\sigma)=\mathrm{Tr}\sqrt{\rho}\sqrt{\sigma}~.
\end{equation}
Up to a factor $2$, these terms provide the (single-shot) formulae for the
quantum Chernoff and Battacharrya bounds, which are used to estimate the
minimum error probability in the discrimination of $\rho$ and $\sigma$ via a
single quantum measurement, i.e., we have%
\begin{equation}
P_{err}\leq\frac{C}{2}\text{~,~}P_{err}\leq\frac{B}{2}~.
\end{equation}
It is straightforward to prove the following chain of inequalities involving
the fidelity%
\begin{equation}
C\leq B\leq\sqrt{F}~. \label{ChainIN}%
\end{equation}
In fact $C\leq B$ is trivial, while $B\leq\sqrt{F}$ comes from the fact
that~\cite{Jozsa,Fuchs}%
\begin{equation}
\mathrm{Tr}\sqrt{\rho}\sqrt{\sigma}=\left\vert \mathrm{Tr}\sqrt{\rho}%
\sqrt{\sigma}\right\vert \leq\mathrm{Tr}\left\vert \sqrt{\rho}\sqrt{\sigma
}\right\vert =\mathrm{Tr}\sqrt{\sqrt{\rho}\sigma\sqrt{\rho}}~,
\end{equation}
where we exploit the inequality $\left\vert \mathrm{Tr}O\right\vert
\leq\mathrm{Tr}\left\vert O\right\vert $ valid for any trace-class operator
$O$.

\section{Quantum fidelity between a pure state and a mixed state\label{Sec2}}

In this section we focus our attention to the case of a mixed state $\rho$ and
a pure state $\sigma=\left\vert \varphi\right\rangle \left\langle
\varphi\right\vert $. In this specific case, we prove that the fidelity can be
simply expressed as a limit formula involving the $s-$overlap.

Before stating this result, it is important to note that the Chernoff term of
Eq.~(\ref{C_inf_rho_sigma}) is defined in terms of an \textit{infimum} over
the \textit{open} interval $(0,1)$. In fact, despite the $s-$overlap
$C_{s}(\rho,\sigma)\leq1$ is correctly defined for any $s$ in the closed
interval $[0,1]$, the two border points $s=0$ and $s=1$ can be excluded from
its minimization, since we always have
\[
C_{0}=\mathrm{Tr}\left\vert \varphi\right\rangle \left\langle \varphi
\right\vert =1~,~C_{1}=\mathrm{Tr}\rho=1~.
\]
Besides the restriction of the interval $[0,1]\rightarrow(0,1)$, it is also
essential to consider an infimum instead of a minimum in
Eq.~(\ref{C_inf_rho_sigma}). In fact, there are nontrivial situations where a
minimum does not exist and an infimum is defined in the limit of
$s\rightarrow0^{+}$ or $s\rightarrow1^{-}$. This is exactly what happens when
one of the two states is pure. In this case the $s-$overlap $C_{s}$ tends to
the quantum fidelity, which becomes equal to the Chernoff term. These are the
main contents of the following results.

\begin{theorem}
\label{THEO}Given a mixed state $\rho$ and a pure state $\left\vert
\varphi\right\rangle \left\langle \varphi\right\vert $, their quantum fidelity
can be expressed as
\begin{equation}
F(\rho,\left\vert \varphi\right\rangle )=\lim_{s\rightarrow1^{-}}C_{s}%
(\rho,\left\vert \varphi\right\rangle )=\lim_{s\rightarrow1^{-}}%
\mathrm{Tr}\left(  \rho^{s}\left\vert \varphi\right\rangle \left\langle
\varphi\right\vert ^{1-s}\right)  ~. \label{Theo1}%
\end{equation}

\end{theorem}

\noindent\textbf{Proof.}~Specify the definition of Eq.~(\ref{Cs_AUX}) to the
case where $\sigma=\left\vert \varphi\right\rangle \left\langle \varphi
\right\vert $, i.e.,%
\begin{equation}
C_{s}=\mathrm{Tr}\left(  \rho^{s}\left\vert \varphi\right\rangle \left\langle
\varphi\right\vert ^{1-s}\right)  ~.\label{Eq_infimum}%
\end{equation}
For every $s\in(0,1)$ we can use the property of the projector
\begin{equation}
\left\vert \varphi\right\rangle \left\langle \varphi\right\vert ^{1-s}%
=\left\vert \varphi\right\rangle \left\langle \varphi\right\vert ~,
\end{equation}
and write%
\begin{equation}
C_{s}=\left\langle \varphi\right\vert \rho^{s}\left\vert \varphi\right\rangle
~.\label{Cs_proof_PURE}%
\end{equation}
Now, we can always decompose $\rho$ as%
\begin{equation}
\rho=\sum_{k}p_{k}^{1/2}\left\vert k\right\rangle \left\langle k\right\vert
~,\label{Rho_zero_expansion}%
\end{equation}
where $p_{k}\in\lbrack0,1]$ for any $k$, and $\{\left\vert k\right\rangle \}$
is an orthonormal set (this is just the spectral decomposition of the state).

Taking the $s$-power of Eq.~(\ref{Rho_zero_expansion}), we get%
\begin{equation}
\rho^{s}=\sum_{k}p_{k}^{s/2}\left\vert k\right\rangle \left\langle
k\right\vert ~,
\end{equation}
for any $s\in(0,1)$. Using the latter equation in Eq.~(\ref{Cs_proof_PURE}),
we achieve%
\begin{equation}
C_{s}=\sum_{k}p_{k}^{s/2}\left\vert \left\langle k\right\vert \left.
\varphi\right\rangle \right\vert ^{2}~.\label{CsnonIN}%
\end{equation}
Finally, taking the limit of $s\rightarrow1^{-}$, we derive%
\begin{align}
\lim_{s\rightarrow1^{-}}C_{s}  & =\sum_{k}p_{k}^{1/2}\left\vert \left\langle
k\right\vert \left.  \varphi\right\rangle \right\vert ^{2}\nonumber\\
& =\left\langle \varphi\right\vert \rho\left\vert \varphi\right\rangle
=F(\rho,\left\vert \varphi\right\rangle )~,\label{End_Proof_Cs}%
\end{align}
which corresponds to the result of Eq.~(\ref{Theo1}).~$\blacksquare$

Note that we can equivalently write%
\begin{equation}
F(\left\vert \varphi\right\rangle ,\rho)=\lim_{s\rightarrow0^{+}}%
C_{s}(\left\vert \varphi\right\rangle ,\rho)=\lim_{s\rightarrow0^{+}%
}\mathrm{Tr}\left(  \left\vert \varphi\right\rangle \left\langle
\varphi\right\vert ^{s}\rho^{1-s}\right)  ~.
\end{equation}
As an application of the previous theorem, we have the following corollary,
which is a result already known in the literature (e.g., see
Refs.~\cite{Kargin,QCbound}).

\begin{corollary}
Given a mixed state $\rho$ and a pure state $\left\vert \varphi\right\rangle
\left\langle \varphi\right\vert $, their quantum fidelity can be expressed as
\begin{equation}
F(\rho,\left\vert \varphi\right\rangle )=C(\rho,\left\vert \varphi
\right\rangle )~.
\end{equation}

\end{corollary}

\noindent\textbf{Proof.}~This is a trivial consequence of the previous
theorem. In fact, we have that $C_{s}$ in Eq.~(\ref{CsnonIN}) is manifestly
non-increasing in $s$. As a consequence, we have%
\begin{equation}
C=\underset{s\in(0,1)}{\mathrm{\inf}}C_{s}=\lim_{s\rightarrow1^{-}}%
C_{s}=F(\rho,\left\vert \varphi\right\rangle )~,
\end{equation}
which completes the proof.~$\blacksquare$

The limit formula of Theorem~\ref{THEO} is useful in all those scenarios where
the $s-$overlap $C_{s}$ is easy to compute. For instance, one of these
scenarios is that of Gaussian states. As we show in the next section, we can
derive a very simple formula for the fidelity between a pure and a mixed
Gaussian state in terms of their statistical moments.

\section{Formula for Gaussian states\label{Sec3}}

In this section we apply the limit formula to the case of multimode Gaussian
states. We first review some basic facts about bosonic systems, symplectic
algebra and Gaussian states. Then, we introduce the notion of symplectic
action, that we use to re-formulate the expression of the $s-$overlap between
two arbitrary Gaussian states. From this expression, we finally derive the
formula for the fidelity between two multimode Gaussian states, in the case
where one of the two states is pure.

\subsection{Basic notions about Gaussian states}

Let us consider a bosonic system of $n$ modes. This quantum system is
described by a tensor product Hilbert space $\mathcal{H}^{\otimes n}$ and a
vector of quadrature operators
\begin{equation}
\mathbf{\hat{x}}^{T}:=(\hat{q}_{1},\hat{p}_{1},\ldots,\hat{q}_{n},\hat{p}%
_{n})~,
\end{equation}
satisfying the commutation relations~\cite{notationCOMM}%
\begin{equation}
\lbrack\mathbf{\hat{x}},\mathbf{\hat{x}}^{T}]=2i\mathbf{\Omega}~,
\label{CommQUAD}%
\end{equation}
where%
\begin{equation}
\mathbf{\Omega}:=\bigoplus\limits_{i=1}^{n}\left(
\begin{array}
[c]{cc}%
0 & 1\\
-1 & 0
\end{array}
\right)  ~. \label{Symplectic_Form}%
\end{equation}
The matrix of Eq.~(\ref{Symplectic_Form}) defines a symplectic form in
$\mathbb{R}^{2n}$. Correspondingly, a real matrix $\mathbf{S}$ is called
\textquotedblleft symplectic\textquotedblright\ when it preserves
$\mathbf{\Omega}$ by congruence, i.e.,%
\begin{equation}
\mathbf{S\Omega S}^{T}=\mathbf{\Omega}~.
\end{equation}

By definition a quantum state $\rho$ of a bosonic system is called
\textquotedblleft Gaussian\textquotedblright\ when its phase-space
representation is Gaussian~\cite{RMP}. In such a case, the quantum state is
completely described by the first two statistical moments. Thus, a Gaussian
state $\rho$ of $n$ bosonic modes is characterized by a displacement vector
\begin{equation}
\mathbf{\bar{x}}:=\mathrm{Tr}(\mathbf{\hat{x}}\rho)~,
\end{equation}
and a covariance matrix (CM)
\begin{equation}
\mathbf{V}:=\tfrac{1}{2}\mathrm{Tr}\left(  \left\{  \mathbf{\hat{x},\hat{x}%
}^{T}\right\}  \rho\right)  -\mathbf{\bar{x}\bar{x}}^{T}~,
\end{equation}
where $\{,\}$ denotes the anticommutator~\cite{notationCOMM}. According to the
definition, a CM\ is a $2n\times2n$ real and symmetric matrix. Furthermore, it
must satisfy the uncertainty principle~\cite{SIMONprinc}%
\begin{equation}
\mathbf{V}+i\mathbf{\Omega}\geq0~. \label{unc_PRINC}%
\end{equation}
A Gaussian state is pure if and only if its CM\ has unit determinant. In fact,
one can easily prove that
\begin{equation}
\mathrm{Tr}\rho^{2}=\frac{1}{\sqrt{\det\mathbf{V}}}~, \label{purity}%
\end{equation}
for a Gaussian state.

\subsection{Symplectic action}

An important tool in the study of Gaussian states is Williamson's
theorem~\cite{Willy}, which assures the symplectic decomposition of a generic
CM. In fact, for every CM\ $\mathbf{V}$, there exists a symplectic matrix
$\mathbf{S}$ such that%
\begin{equation}
\mathbf{V}=\mathbf{SWS}^{T}~,
\end{equation}
where%
\begin{equation}
\mathbf{W}=\bigoplus\limits_{i=1}^{n}\nu_{i}\mathbf{I}\text{~~},~~~\mathbf{I}%
:=\left(
\begin{array}
[c]{cc}%
1 & \\
& 1
\end{array}
\right)  .
\end{equation}
The matrix $\mathbf{W}$\ is called the \textquotedblleft Williamson
form\textquotedblright\ of $\mathbf{V}$, and the set $\{\nu_{i}\}=\{\nu
_{1},\cdots,\nu_{n}\}$ is called the \textquotedblleft symplectic
spectrum\textquotedblright\ of $\mathbf{V}$. As a consequence of the
uncertainty principle, each symplectic eigenvalue $\nu_{i}$ must be greater
than or equal to the quantum shot-noise (here corresponding to $1$). More
exactly, the uncertainty principle of Eq.~(\ref{unc_PRINC}) is equivalent to
the conditions~\cite{RMP,Alex}%
\begin{equation}
\mathbf{V}>0~,~\nu_{i}\geq1~(\text{for any }i)~. \label{Heis2}%
\end{equation}
In particular, a Gaussian state is pure if and only if its symplectic spectrum
is all equal to one ($\nu_{i}=1$ for any $i$). In other words, for a pure
Gaussian state, the Williamson form is equal to the identity. This is a direct
consequence of Eqs.~(\ref{purity}) and~(\ref{Heis2}) plus the fact that the
determinant is a global symplectic invariant (i.e., $\det\mathbf{V}%
=\det\mathbf{W}$).

Now, consider a real function $f:$ $\mathbb{R\rightarrow R}$ and generic CM
$\mathbf{V}$ with symplectic decomposition%
\begin{equation}
\mathbf{V}=\mathbf{S~}\left[  \bigoplus\limits_{i=1}^{n}\nu_{i}\mathbf{I}%
\right]  \mathbf{~S}^{T}~.
\end{equation}
Then, we define the \textquotedblleft symplectic action\textquotedblright%
\ $f(\mathbf{V})_{\ast}$ of $f$ over $\mathbf{V}$ the following matrix%
\begin{equation}
f(\mathbf{V})_{\ast}=\mathbf{S}~\left[  \bigoplus\limits_{i=1}^{n}f(\nu
_{i})\mathbf{I}\right]  ~\mathbf{S}^{T}~.
\end{equation}
Since the symplectic decomposition is unique (up to uninfluential local
rotations), the output matrix $f(\mathbf{V})_{\ast}$ is unambiguously defined.
In particular, this matrix is a CM\ if and only if $f(\nu_{i})\geq1$ for every
$i$. It is also clear that $f(\mathbf{SVS}^{T})_{\ast}=\mathbf{S}%
f(\mathbf{V})_{\ast}\mathbf{S}^{T}$ for every CM\ $\mathbf{V}$\ and symplectic
matrix $\mathbf{S}$.

It is important to note that this operation is different from the standard
notion of function of a matrix $f(\mathbf{V})$, where $f$ is applied to the
standard eigenvalues of the spectral decomposition of $\mathbf{V}$. We have
$f(\mathbf{V})_{\ast}=f(\mathbf{V})$\ only if spectral and symplectic
decompositions coincide, which happens when the symplectic matrix $\mathbf{S}%
$\ is a proper rotation (so that $\mathbf{S}^{T}=\mathbf{S}^{-1}$). In
general, the symplectic action is a useful tool which enables us to simplify
the formalism in the manipulation of the CMs.

\subsection{From the $s-$overlap to the quantum fidelity}

According to Ref.~\cite{MinkoPRA}, we can write a closed formula for the
$s-$overlap between two arbitrary multimode Gaussian states. Here we briefly
review this formula by adopting the formalism of the symplectic action.

First of all, let us define the two real functions%
\begin{equation}
G_{p}(x):=\frac{2^{p}}{\left(  x+1\right)  ^{p}-\left(  x-1\right)  ^{p}}~,
\label{G_function}%
\end{equation}
and%
\begin{equation}
\Lambda_{p}(x):=\frac{\left(  x+1\right)  ^{p}+\left(  x-1\right)  ^{p}%
}{\left(  x+1\right)  ^{p}-\left(  x-1\right)  ^{p}}~,
\end{equation}
which are finite and non-negative for every $x\geq1$ and $p>0$. Using these
functions, we can easily express the $s-$overlap between two arbitrary
$n$-mode Gaussian states, $\rho_{0}$ and $\rho_{1}$, with statistical moments
$\{\mathbf{\bar{x}}_{0},\mathbf{V}_{0}\}$\ and $\{\mathbf{\bar{x}}%
_{1},\mathbf{V}_{1}\}$, and associated symplectic spectra $\{\nu_{i}^{0}\}$
and $\{\nu_{i}^{1}\}$. In fact, for any $0<s<1$, their $s-$overlap is given by%
\begin{equation}
C_{s}(\rho_{0},\rho_{1})=\Pi_{s}\left(  \det\Sigma_{s}\right)  ^{-1/2}%
\exp\left(  -\frac{\mathbf{d}^{T}\Sigma_{s}^{-1}\mathbf{d}}{2}\right)  ~,
\label{Cs_QChernoff}%
\end{equation}
where $\mathbf{d}:=\mathbf{\bar{x}}_{0}-\mathbf{\bar{x}}_{1}$,%
\begin{equation}
\Sigma_{s}:=\Lambda_{s}(\mathbf{V}_{0})_{\ast}+\Lambda_{1-s}(\mathbf{V}%
_{1})_{\ast}~, \label{SIGMAs}%
\end{equation}
and%
\begin{equation}
\Pi_{s}:=2^{n}\prod\limits_{i=1}^{n}G_{s}(\nu_{i}^{0})G_{1-s}(\nu_{i}^{1})~.
\end{equation}
Note that the symplectic action intervenes in Eq.~(\ref{SIGMAs}). Explicitly,
we have%
\begin{align}
\mathbf{V}_{0}  &  =\mathbf{S}_{0}\left[  \bigoplus\limits_{i=1}^{n}\nu
_{i}^{0}\mathbf{I}\right]  \mathbf{S}_{0}^{T}\nonumber\\
&  \rightarrow\Lambda_{s}(\mathbf{V}_{0})_{\ast}=\mathbf{S}_{0}\left[
\bigoplus\limits_{i=1}^{n}\Lambda_{s}(\nu_{i}^{0})\mathbf{I}\right]
\mathbf{S}_{0}^{T}~,
\end{align}
and%
\begin{align}
\mathbf{V}_{1}  &  =\mathbf{S}_{1}\left[  \bigoplus\limits_{i=1}^{n}\nu
_{i}^{1}\mathbf{I}\right]  \mathbf{S}_{1}^{T}\nonumber\\
&  \rightarrow\Lambda_{1-s}(\mathbf{V}_{1})_{\ast}=\mathbf{S}_{1}\left[
\bigoplus\limits_{i=1}^{n}\Lambda_{1-s}(\nu_{i}^{1})\mathbf{I}\right]
\mathbf{S}_{1}^{T}~.
\end{align}

The formula of the $s$-overlap can be greatly simplified in the presence of
pure Gaussian states, on which the two functions $\Lambda_{p}$ and $G_{p}$
have a trivial action. In fact, suppose that a Gaussian state $\rho$ is pure.
This means that its symplectic spectrum is all equal to one, i.e., $\nu_{i}=1$
for any $i$. In other words, its CM has symplectic decomposition%
\begin{equation}
\mathbf{V}=\mathbf{S}\left[  \bigoplus\limits_{i=1}^{n}\mathbf{I}\right]
\mathbf{S}^{T}~,
\end{equation}
where the Williamson form corresponds to the $n$-mode identity matrix. Then,
for every $p>0$, we have%
\begin{equation}
\Lambda_{p}(\mathbf{V})_{\ast}=\mathbf{V~,} \label{Invariance_PURECM}%
\end{equation}
i.e., the symplectic action of $\Lambda_{p}$ does not change pure CMs. In
fact, explicitly we have%
\begin{equation}
\Lambda_{p}(\mathbf{V})_{\ast}=\mathbf{S}\left[  \bigoplus\limits_{i=1}%
^{n}\Lambda_{p}(1)\mathbf{I}\right]  \mathbf{S}^{T}=\mathbf{S}\left[
\bigoplus\limits_{i=1}^{n}\mathbf{I}\right]  \mathbf{S}^{T}=\mathbf{V}~,
\end{equation}
where we use the fact that $\Lambda_{p}(1)=1$ for any $p>0$. Also the
computation of $G_{p}$ becomes trivial. In fact, for any $p>0$ we have%
\begin{equation}
G_{p}(\nu_{i})=G_{p}(1)=1~. \label{GpPURE}%
\end{equation}
Coming back to the formula of Eq.~(\ref{Cs_QChernoff}), if one of the two
Gaussian state is pure, e.g., $\rho_{1}=\left\vert \varphi_{1}\right\rangle
\left\langle \varphi_{1}\right\vert $, then we have the simplifications%
\begin{equation}
\Pi_{s}=2^{n}\prod\limits_{i=1}^{n}G_{s}(\nu_{i}^{0})~, \label{PI_sPROOF}%
\end{equation}
and%
\begin{equation}
\Sigma_{s}=\Lambda_{s}(\mathbf{V}_{0})_{\ast}+\mathbf{V}_{1}~,
\label{SIGMA_sPROOF}%
\end{equation}
for every $s\in(0,1)$. Now, by taking the limit of $s\rightarrow1^{-}$, we can
derive the formula for Gaussian states.

\begin{theorem}
Let us consider two $n$-mode Gaussian states, $\rho_{0}$ and $\rho_{1}$, where
$\rho_{0}$ is generally mixed (with moments $\mathbf{\bar{x}}_{0}$ and
$\mathbf{V}_{0}$) and $\rho_{1}=\left\vert \varphi_{1}\right\rangle
\left\langle \varphi_{1}\right\vert $ is pure (with moments $\mathbf{\bar{x}%
}_{1}$ and $\mathbf{V}_{1}$). Their fidelity $F=F(\rho_{0},\left\vert
\varphi_{1}\right\rangle )$ can be computed via the formula%
\begin{equation}
F=\frac{2^{n}}{\sqrt{\det(\mathbf{V}_{0}+\mathbf{V}_{1})}}\exp\left[
-\frac{\mathbf{d}^{T}\left(  \mathbf{V}_{0}+\mathbf{V}_{1}\right)
^{-1}\mathbf{d}}{2}\right]  , \label{Fid_FormulaNEW}%
\end{equation}
where $\mathbf{d}:=\mathbf{\bar{x}}_{0}-\mathbf{\bar{x}}_{1}$.
\end{theorem}

\noindent\textbf{Proof.}~This proof simply combines the limit formula of the
fidelity, given in Eq.~(\ref{Theo1}), with the analytical formula of $C_{s}$
for Gaussian states. Given two Gaussian states $\rho_{0}$ and $\rho
_{1}=\left\vert \varphi_{1}\right\rangle \left\langle \varphi_{1}\right\vert
$, their $s-$overlap $C_{s}(\rho_{0},\left\vert \varphi_{1}\right\rangle )$ is
expressed by Eq.~(\ref{Cs_QChernoff}) with the simplified terms of
Eqs.~(\ref{PI_sPROOF}) and~(\ref{SIGMA_sPROOF}), where $\{\nu_{i}^{0}\}$ is
the symplectic spectrum of $\rho_{0}$. Now taking the limit of $s\rightarrow
1^{-}$ in $C_{s}(\rho_{0},\left\vert \varphi_{1}\right\rangle )$ is equivalent
to taking the limit of $s\rightarrow1^{-}$ in the terms $\Pi_{s}$ and
$\Sigma_{s}$ of Eqs.~(\ref{PI_sPROOF}) and~(\ref{SIGMA_sPROOF}).

Since $G_{s}$ and $\Lambda_{s}$ are continuous at $s=1$, we have%
\begin{align}
\lim_{s\rightarrow1^{-}}G_{s}(x)  &  =G_{1}(x)=1~,\\
\lim_{s\rightarrow1^{-}}\Lambda_{s}(x)  &  =\Lambda_{1}(x)=x~,
\end{align}
for every $x\geq1$. In particular, for every CM $\mathbf{V}$, we can write%
\begin{equation}
\lim_{s\rightarrow1^{-}}\Lambda_{s}(\mathbf{V})_{\ast}=\Lambda_{1}%
(\mathbf{V})_{\ast}=\mathbf{V}~.
\end{equation}
By applying these properties to the state $\rho_{0}$, we get%
\begin{equation}
\lim_{s\rightarrow1^{-}}G_{s}(\nu_{i}^{0})=1~,~\lim_{s\rightarrow1^{-}}%
\Lambda_{s}(\mathbf{V}_{0})_{\ast}=\mathbf{V}_{0}~.
\end{equation}
As a consequence, we can write
\begin{equation}
\lim_{s\rightarrow1^{-}}\Pi_{s}=2^{n}~,~\lim_{s\rightarrow1^{-}}\Sigma
_{s}=\mathbf{V}_{0}+\mathbf{V}_{1}~. \label{PI_SIGMA_limit}%
\end{equation}
Now, using Eq.~(\ref{PI_SIGMA_limit}) and Eq.~(\ref{Cs_QChernoff}), we get%
\begin{equation}
\lim_{s\rightarrow1^{-}}C_{s}(\rho_{0},\left\vert \varphi_{1}\right\rangle
)=\frac{2^{n}}{\sqrt{\det(\mathbf{V}_{0}+\mathbf{V}_{1})}}e^{-\frac
{\mathbf{d}^{T}\left(  \mathbf{V}_{0}+\mathbf{V}_{1}\right)  ^{-1}\mathbf{d}%
}{2}},
\end{equation}
which provides the result of the theorem.~$\blacksquare$

\section{Conclusion\label{Sec4}}

In conclusion, we have provided a limit formula for computing the quantum
fidelity between a mixed and a pure state. This formula involves a generalized
$s$-overlap between the two quantum states, a quantity used in the definition
of the quantum Chernoff bound. As an application of the formula, we have
considered the case of Gaussian states, for which we have derived a simple
expression in terms of their first and second-order statistical moments.

An alternative formula for the computation of the quantum fidelity can be
useful in many scenarios, including protocols of quantum
teleportation~\cite{CVtelepo,Bra98,RalphTELE,TeleNET,PirTeleOPMecc,Barlett2003,Pirgames,Pir2005,teleREV,Sherson}%
, entanglement swapping~\cite{Entswap,Entswap2,PirENTswap}, and quantum
cloning~\cite{Buzek,lin,Cerf1,braclone,fiu}. Clearly, other important areas of
application are quantum state discrimination (i.e., quantum hypothesis
testing) and quantum channel discrimination, where the latter includes
practical problems such as the quantum illumination of
targets~\cite{QIll1,QIll2} and the quantum reading of classical digital
memories~\cite{QreadingPRL,QreadCAP,Nair11,Hirota11,Bisio11,Arno11,Saikat}.

\section*{Acknowledgments}
C.W. acknowledges support from the Ontario postdoctoral fellowship
program, the CQIQC postdoctoral fellowship program, CIFAR, the
Canada Research Chair program, NSERC, QuantumWorks. S. P.
acknowledges support from EPSRC (Grant No. EP/J00796X/1). Authors
would like to thank Aharon Brodutch for spotting a previous error
in the proof of Theorem 1.


\begin{thebibliography}{99}
\bibitem {BuresFID}An alternative definition which is commonly adopted in
quantum information is the square-root or Bures' fidelity. This corresponds to
the square root of the Uhlmann-Jozsa fidelity%
\[
F_{B}(\rho,\sigma):=\sqrt{F(\rho,\sigma)}=\mathrm{Tr}\sqrt{\sqrt{\rho}%
\sigma\sqrt{\rho}}~.
\]
Bures' fidelity is the direct generalization of the classical fidelity. In
fact, if the two density operators commute, we can write $\rho=\sum_{k}%
p_{k}\left\vert k\right\rangle \left\langle k\right\vert $ and $\sigma
=\sum_{k}q_{k}\left\vert k\right\rangle \left\langle k\right\vert $ for an
orthonormal basis $\{\left\vert k\right\rangle \}$. Then, we have $F_{B}%
(\rho,\sigma)=\sum_{k}\sqrt{p_{k}q_{k}}$ which is the classical fidelity
$F(p_{k},q_{k})$ between the two probability distributions $p_{k}$ and $q_{k}$.
\bibitem {Kolmo}This is the quantum version of the Kolmogorov distance
$D(p_{k},q_{k}):=\frac{1}{2}\sum_{k}\left\vert p_{k}-q_{k}\right\vert $ which
is defined for two classical probability distributions $\{p_{k}\}$ and
$\{q_{k}\}$.
\bibitem {TraceCLASS}Roughly speaking, a trace-class operator is a linear
operator for which we can define the trace. More rigorously, a bounded linear
operator $O$\ over a separable Hilbert space $\mathcal{H}$\ is called
\textquotedblleft trace-class\textquotedblright\ if for some (and hence all)
orthonormal bases $\left\{  \left\vert \varphi_{k}\right\rangle \right\}  $ we
have $\sum_{k}\left\langle \varphi_{k}\right\vert \sqrt{O^{\dagger}%
O}\left\vert \varphi_{k}\right\rangle <\infty$. In such a case, both the trace
$\mathrm{Tr}O$ and the trace norm $\left\Vert O\right\Vert _{1}:=\mathrm{Tr}%
\sqrt{O^{\dagger}O}$ are correctly defined and finite. Furthermore, we can
define a whole class of Schatten $p$-norms $\left\Vert O\right\Vert
_{p}:=\left(  \mathrm{Tr}\left\vert O\right\vert ^{p}\right)  ^{1/p}$ for
every real $p\geq1$. This class includes the trace norm for $p=1$ and the
Hilbert-Schmidt norm for $p=2$.
\bibitem {notationCOMM}With the notation $[\mathbf{\hat{x}},\mathbf{\hat{x}%
}^{T}]$ we mean a matrix with entries $[\mathbf{\hat{x}}_{i},\mathbf{\hat{x}%
}_{j}]$, where $i,j=1,\cdots,2n$. Analogously, $\{\mathbf{\hat{x}%
},\mathbf{\hat{x}}^{T}\}$ is a matrix with entries $\{\mathbf{\hat{x}}%
_{i},\mathbf{\hat{x}}_{j}\}$.
\end{thebibliography}
\end{document}